\documentclass[showpacs,eqsecnum,aps,nofootinbib,preprintnumbers,floatfix]{revtex4}

\usepackage{epsfig}
\usepackage{citesort}
\usepackage{amssymb}
\usepackage{latexsym}

\def\ds{\displaystyle}
\def\beq{\begin{equation}}
\def\eeq{\end{equation}}
\def\bea{\begin{eqnarray}}
\def\eea{\end{eqnarray}}
\def\beeq{\begin{eqnarray}}
\def\eeeq{\end{eqnarray}}

\def\etal{{\it et al.}}

\begin{document}

\title{Analysis of the $B \to K^*_{2} (\to K \pi) l^+ l^-$ decay}

  \author{S. Rai Choudhury}
     \email{src@jamia-physics.net}
     \affiliation{Centre for Theoretical Physics, Jamia Millia, New Delhi 110 025, India}
  \author{A. S. Cornell}
   \email{alanc@yukawa.kyoto-u.ac.jp}
   \affiliation{Yukawa Institute for Theoretical Physics, Kyoto University, Kyoto 606-8502, Japan}
   \author{G. C. Joshi}
     \email{joshi@physics.unimelb.edu.au}
     \affiliation{School of Physics, University of Melbourne, Victoria 3010, Australia}
   \author{B. H. J. McKellar}
     \email{b.mckellar@physics.unimelb.edu.au}
     \affiliation{School of Physics, University of Melbourne, Victoria 3010, Australia}
   
\date{\today}


%%%%%%%%%%%%%%%

\begin{abstract}
In this paper we study the angular distribution of the rare $B$-decay $B \to K^*_2 (\to K \pi) l^+ l^-$, which is expected to be observed soon. We use the standard effective Hamiltonian approach, and use the form factors that have already been estimated for the corresponding radiative decay $B \to K^*_2 \gamma$. The additional form factors that come into play for the dileptonic channel are estimated using  the large energy effective theory (LEET), which enables one to relate the additional form factors to the form factors for the radiative mode. Our results provide, just like in the case of the $K^*(892)$ resonance, an opportunity for a straightforward comparison of the basic theory with experimental results, which may be expected in the near future for this channel.
\end{abstract}

\maketitle

%%%%%%%%%%%%%%%%%%%%%%%%%%%%%%%%%%%%%%%%%%%%%%%%%%%%%%%%%%%%%%%%%%%%%
%
%  Section 1: Introduction
%
%%%%%%%%%%%%%%%%%%%%%%%%%%%%%%%%%%%%%%%%%%%%%%%%%%%%%%%%%%%%%%%%%%%%%

\section{Introduction}

\par The flavour changing neutral current (FCNC) decays, like $b \to s l^+ l^-$ or $b \to d l^+ l^-$, have attracted a lot of attention during the last decade \cite{Buchalla:1995vs,Greub:1999sv,Ali:2005}, this being due to the fact that these processes are forbidden at the tree level in the standard model (SM) but are induced by loop corrections. They are therefore very sensitive to details of the model and hence form a very appropriate process for the study of physics of the SM and beyond. Using the standard theoretical framework for the study of these processes, this framework being based on the short distance expansion of the relevant quark operators \cite{Buchalla:1995vs}, the values of the Wilson Coefficients that enter the expansion have now been calculated to a very high degree of accuracy. As such the major uncertainty now lies in the evaluation of the matrix elements of the relevant quark operators between the physical states of the hadrons in the process being considered. 

\par For semileptonic decays, like  $B \to K^* l^+ l^-$, the relevant quark operators are bilinear in the quark fields, and the matrix elements of the quark operators reduce to form factors.  The form factors involved in this case are those of the vector, axial vector and tensor current operators between the initial $B$ and final $K^*$ meson. There have been many theoretical analyses of these form factors and a most promising approach for our purposes is the large energy effective theory (LEET) \cite{Dugan:1990de}. The LEET theory brings in an enormous economy of parameters since it yields relations between the various form factors involved \cite{Ali:2002qc}. Once the form factors are estimated, the formalism permits the calculation of the full amplitude for the decay in all helicity states. Note that there have been calculations before of lepton polarization for these decays \cite{Choudhury:2003mi}. For the $K^*(892)$ case, a calculation by Kim \etal\cite{Kim:2000dq} pointed out the possibility of studying the azimuthal angular distribution of the decay products of the $K^*$ as a probe for new physics. Of late, the Babar group \cite{Aubert:2003zs} has obtained experimental estimates of the radiative decays of the decay $B \to K^*_2(1430) \gamma$. The decay rates are comparable with the corresponding one for the $K^*(892)$. We can therefore hope that in analogy to the $K^*(892)$ case, data on decays of $B \to  K^*_2(1430) l^+ l^-$ would also be availabe in the near future. We carry out in this work an analysis of azimuthal angular dependence of the decay products of the $K^*$ resonance similar to the analysis done by Kim \etal\cite{Kim:2000dq}. The radiative decay for this resonance has already been analyzed by Cheng \etal\cite{Cheng:2004yj} and there is reasonable agreement between theory and experiment for this process. Some, but not all, of the form factors involved in the corresponding dileptonic mode are related to the radiative mode. However, the LEET relates the form factors of various quark bilinears, and these relations suffice to tackle the dileptonic mode without introducing any new parameters. The angular distributions involve the full density matrix rather than simply the diagonal ones for a decay rate, and thus would prove to be an even better test of the underlying theory than the decay rates of the radiative mode alone.

%%%%%%%%%%%%%%%%%%%%%%%%%%%%%%%%%%%%%%%%%%%%%%%%%%%%%%%%%%%%%%%%%%%%%
%
%  Section 2: The Equations
%
%%%%%%%%%%%%%%%%%%%%%%%%%%%%%%%%%%%%%%%%%%%%%%%%%%%%%%%%%%%%%%%%%%%%%

\section{The Equations}

\begin{figure}
\begin{center}
\epsfig{file=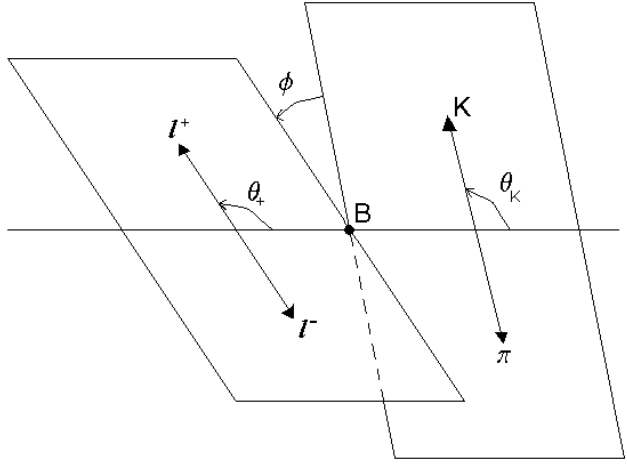,angle=90,width=.5\textwidth}
\caption{\it A pictorial representation of the process $B \to K^*_2 (\to K \pi) l^+ l^-$.}  
\label{fig:l}
\end{center}
\end{figure}

\par The full process, together with a representation of the kinematics we have used, is shown pictorially in figure 1. This figure also shows the labelling of the momenta of the various particles involved in this process. Following now the standard effective Hamiltonian approach, we first note that the $B \to K^* l^+ l^-$ process is described at the quark level by the process 
$b \to s l^+ l^-$. In the limit where we neglect the mass of the strange quark, this is described by the effective Hamiltonian:
\bea
H_{eff} & = & \left(\frac{G_F \alpha}{\sqrt{2} \pi}  \lambda_{CKM}\right) \Big[(C_9 -C_{10}) (\bar s \gamma_{\mu} b_L) (\bar l_L \gamma^{\mu} l_L ) + ( C_9 + C_{10} ) (\bar s \gamma_{\mu} b_L ) ( \bar l_R \gamma^{\mu} l_R) \nonumber \\
&& \hspace{3cm} - 2 C_{7L} \left( \bar s i \sigma_{\mu\nu} \frac{ m_b q^{\nu}}{q^2} b \right) \big[(\bar l_L \gamma^{\mu} l_L) + ( l_R \gamma^{\mu} l_R ) \big] \Big] ~,~
\eea
where $q = (p_+ + p_-)$. As such, for the process under consideration, we need an estimation of the matrix element for the vector, axial vector, tensor and the pseudotensor bilinears between the $\vert B \rangle$ and $\vert K^*_2 \rangle$ states. These can be expressed in terms of invariant form factors as follows:
\bea
i \langle K^*_2(p_{K^*}) \vert \bar s \sigma_{\mu \nu} q^{\nu} \gamma_5 b \vert B(p_B)\rangle & = & \frac{1}{m_B} (p_B + p_{K^*})^{\sigma} \Big[ \epsilon^*_{\mu\sigma}(p_{K^*} + p_B).q - (p_{K^*} + p_B)_{\mu} \epsilon^*_{\sigma \rho} q^{\rho} \Big] U_2  \nonumber \\
&& \hspace{1cm} - \epsilon^*_{\sigma \rho} ( p_B + p_{K^*})^{\sigma} q^{\rho} \Big[ q_{\mu} - 
(p_{K^*} + p_B)_{\mu} \frac{q^2}{(p_{K^*} + p_B).q} \Big] \frac{U_3}{m_B}
\eea
where $q^2= ( p_B - p_{K^*})^2$, $q.(p_B + p_{K^*}) = m_B^2 - m_{K^*}^2$, $\epsilon^*_{\sigma \rho}. p_{K^*}^{\sigma} = 0$, and:
\bea
i \langle K^*_2(p_{K^*}) \vert \bar s \sigma_{\mu \nu} q^{\nu} b \vert B(p_B)\rangle = \frac{ 2 i}{m_B}  U_1  \epsilon_{\mu \nu \lambda \rho} \epsilon^{* \nu \sigma} p_{B \sigma} p_B^{\lambda} p_{K^*}^{\rho} ~,~
\eea 
\beq
\langle K^*_2(p_{K^*}) \vert \bar s \gamma_{\mu} \gamma_5 b \vert B(p_B) \rangle = \epsilon^{* \alpha \beta } \Big [A_1 \left( g_{\alpha \mu} p_{B \beta} + g_{\beta \mu} p_{B \alpha}\right) + 
A_2 ( p_{B\alpha \beta \mu}) + A_3 ( p_{B \alpha}p_{B \beta}p_{K^* \mu})\Big] ~,~
\eeq
and finally:
\beq
\langle K^*_2(p_{K^*})\vert \bar s \gamma_{\mu} b \vert B(p_B)\rangle = \epsilon^{\alpha \beta} \Big[ \epsilon_{\alpha \mu\nu\rho} p_*^{\nu} p_B^{\rho}p_{B \beta} + 
\alpha \leftrightarrow \beta \Big]  V ~.~
\eeq 
Note that the form factors $U_1$, $U_2$ and $U_3$ have been estimated in the covariant light cone approach by Cheng and Chua \cite{Cheng:2004yj}, where their results are consistent with the data availabe for the radiative decay of the $B \to K^*_2 \gamma$ process\cite{Aubert:2003zs}. For the estimation of the form factors $A_1$, $A_2$, $A_3$ and $V$ we follow the LEET effective theory analysis given by Charles \etal\cite{Dugan:1990de}. As such, we obtain the relations:
\bea
A_1 & = & - U_2 /2 ~,~ \nonumber \\
A_2 & = & 0 ~,~ \nonumber \\
A_3 & = & 2 \Big[ \frac{U_2}{m_B E} - \frac{U_3}{m_B^3} \Big] ~,~ \nonumber \\
V & = & i \frac{U_1}{m_B^2} ~,~
\eea
where we have taken the limit of the heavy quark mass going to infinity and $E = p_B.p_{K^*}/m_B$. Note that with this approach we have introduced no extra hadronic form factors beyond what is required for the radiative mode. Thus, once we are able to describe the radiative mode we have in effect a check on the model from the dileptonic mode.

\par The final piece required for the study of the full decay process, $B \to K^*_2 (\to K \pi) l^+ l^-$, involves the decay of the $K^*_2 \to K \pi$. Note that the data for this process already exists, where the vertex is described in terms of the coupling constant $g$ by:
\beq
\langle K_(p_K) \pi(p_{\pi})\vert K^*_2(p_{K^*})\rangle= \frac{i g}{m_{K^*}} \epsilon_{\alpha \beta} p_K^{\alpha} p_K^{\beta} ~.~
\eeq
The known width of the $K^*_2$, $\Gamma$, and the branching ratio into this channel can then be used to determine the constant $g$. The amplitude for the process, as given in figure 1, can be written as:
\beq
{\cal M} = {\cal M_L} + {\cal M_R} ~,~
\eeq
with
\bea
{\cal M_L} & = & i \frac{G_F \alpha \lambda_{CKM}}{\sqrt 2 \pi} (\bar l_L 
\gamma^{\mu} l_L )  L_{\mu \alpha \beta} \frac{ i B^{\alpha \beta, \rho \sigma}}{p_{K^*}^2 - m_{K^*}^2 + i m_{K^*} \Gamma} \frac{ i g p_{K \rho} p_{K \sigma} }{m_{K^*}} ~,~ \\
{\cal M_R} & = & i \frac{G_F \alpha \lambda_{CKM}}{\sqrt 2 \pi} (\bar l_R \gamma^{\mu} l_R )  R_{\mu \alpha \beta} \frac{i B^{\alpha \beta, \rho \sigma}}{p_{K^*}^2 - m_{K^*}^2 + i m_{K^*} \Gamma}
\frac{igp_{K \rho}p_{K\sigma}}{m_{K^*}} ~,~
\eea
where
\bea
R_{\mu \alpha \beta} & = & \frac{1}{2} ( C_9 + C_{10} ) \Big[ 2 V \epsilon_{\alpha 
\mu \nu \rho} p_K^{\nu} p_B^{\rho} p_{B \beta} - 2 A_1 g_{\alpha \mu} p_{B \beta} - A_3 p_{B \alpha} p_{B \beta} p_{K \mu} \Big] \nonumber \\
&& \hspace{1cm} - C_{7L} \frac{m_B q^{\nu}}{q^2} \Big[ - U_2 \left( m_B^2 g_{\mu \alpha} 
p_{B \beta}  -(p_K+p_B)_{\mu} \frac{p_{B \alpha} p_{B \beta}}{m_B} \right) \nonumber \\
&& \hspace{1.5cm} + \frac{2}{m_B} U_3 p_{B \alpha}p_{B \beta} \left(p_{K \mu} - (p_K + p_B)_{\mu}\frac{p_k.p_B}{m_B^2} \right) + \frac{2 i U_1}{m_B} \epsilon_{\mu \alpha \lambda \rho} p_B^{\lambda} p_K^{\rho} p_{B \beta} \Big] ~,~ \\
L_{\mu \alpha \beta} & = & R_{\mu \alpha \beta}\left( C_{10} \to - C_{10} \right) ~.~ 
\eea
Where we shall neglect the masses of the lepton, the $K$ and the pion in comparison with the mass of the $B$-meson. Note that this is consistent with our determination of the relations between the form factors, as we have already taken the heavy quark limit. 

\par Using the expression for the amplitude ${\cal M}$ above, we can write: 
\beq
\left| {\cal M} \right|^2 = \left| \ds \frac{\alpha G_F \lambda_{CKM} g}{\sqrt{2} \pi m_{K^*}} \frac{1}{p_{K^*}^2 - m_{K^*}^2 + i m_{K^*} \Gamma_{K^*} } \right|^2 \left( 2 \mathrm{Re} [ {\cal A}_L {\cal A}_R^* ] + |{\cal A}_L|^2 + |{\cal A}_R|^2 \right) ~,~
\eeq
where $\ds \Gamma_{K^*} = \frac{g^2 \lambda^{5/2}}{960 m_{K^*}^9 \pi}$, and where:
\bea
2 \mathrm{Re} [{\cal A}_L {\cal A}_R^*] & = & 0 ~,~ \nonumber
\eea
\bea
\left| {\cal A}_L \right|^2 & = & 4 \epsilon^{\alpha \beta \mu \nu} p_{- \alpha} p_{+ \beta} p_{K \mu} p_{K^* \nu} \left\{ \mathrm{Im} [B C^* + B D^*] + q . p_{K} \mathrm{Re} [A B^*] + q . p_{K^*} \mathrm{Re} [A C^* + A D^*] \right\} \nonumber \\
&& \hspace{0.5cm} + 4 \left( p_- . p_{K^*} p_+ . p_K + p_- . p_K p_+ . p_{K^*} - p_- . p_+ p_K . p_{K^*} \right) \mathrm{Re} [B C^* + B D^*] \nonumber \\
&& \hspace{0.5cm} + 4 \left( - m_{K^*}^2 p_- . p_+ + 2 p_- . p_{K^*} p_+ . p_{K^*} \right) \left( \mathrm{Re} [C D^*] + \frac{1}{2} |D|^2 \right) \nonumber \\
&& \hspace{0.5cm} + 2 \left( 2 p_- . p_K p_+ . p_{K} \right) |B|^2 + 2 \left( 2 p_- . p_{K^*} p_+ . p_{K^*} - m_{K^*}^2 p_- . p_+ \right) |C|^2 \nonumber \\
&& \hspace{0.5cm} + 4 \mathrm{Im} [D A^* + C A^*] \Bigg[ p_+ . p_K (p_- . p_{K^*})^2 - p_- . p_K (p_+ . p_{K^*})^2 - p_- . p_+ p_K . p_{K^*} (p_- - p_+) . p_{K^*} \nonumber \\
&& \hspace{4.5cm} + (p_+ - p_-) . p_K \left( - m_{K^*}^2 p_- . p_+ + p_- . p_{K^*} p_+ . p_{K^*} \right) \Bigg] \nonumber \\
&& \hspace{0.5cm} + 4 \mathrm{Im} [A B^*] \Bigg[ p_+ . p_{K^*} (p_- . p_{K})^2 - p_- . p_{K^*} (p_+ . p_{K})^2 - p_- . p_+ p_K . p_{K^*} (p_- - p_+) . p_{K} \nonumber \\
&& \hspace{3.0cm} + (p_+ - p_-) . p_{K^*} \left( + p_- . p_{K} p_+ . p_{K} \right) \Bigg]
\nonumber \\
&& \hspace{0.5cm} + 4 |A|^2 \Bigg[ - 2 p_- . p_K p_- . p_{K^*} p_+ . p_K p_+ . p_{K^*} + p_+ . p_- p_K . p_{K^*} (p_- - p_+) . p_K (p_- - p_+) . p_{K^*} \nonumber \\
&& \hspace{2.5cm} + (p_- . p_{K^*})^2 (p_+ . p_K)^2 + (p_+ . p_{K^*})^2 (p_- . p_K)^2 + m_{K^*}^2 p_- . p_K p_+ . p_K p_+ . p_- \nonumber \\
&& \hspace{2.5cm} - \left(\frac{1}{2} p_- . p_+\right)\left( m_{K^*}^2 (p_- . p_K)^2 + m_{K^*}^2 (p_+ . p_K)^2 \right) \Bigg] ~,~ 
\eea
\bea
\left| {\cal A}_R \right|^2 & = & 4 \epsilon^{\alpha \beta \mu \nu} p_{- \alpha} p_{+ \beta} p_{K \mu} p_{K^* \nu} \left\{ - \mathrm{Im} [N G^* + N H^*] + q . p_{K} \mathrm{Re} [Q N^*] + q . p_{K^*} \mathrm{Re} [Q G^* + Q H^*] \right\} \nonumber \\
&& \hspace{0.5cm} + 4 \left( p_- . p_{K^*} p_+ . p_K + p_- . p_K p_+ . p_{K^*} - p_- . p_+ p_K . p_{K^*} \right) \mathrm{Re} [N G^* + N H^*] \nonumber \\
&& \hspace{0.5cm} + 4 \left( - m_{K^*}^2 p_- . p_+ + 2 p_- . p_{K^*} p_+ . p_{K^*} \right) \left( \mathrm{Re} [G H^*] + \frac{1}{2} |H|^2 \right) \nonumber \\
&& \hspace{0.5cm} + 2 \left( 2 p_- . p_K p_+ . p_{K} \right) |N|^2 + 2 \left( 2 p_- . p_{K^*} p_+ . p_{K^*} - m_{K^*}^2 p_- . p_+ \right) |G|^2 \nonumber \\
&& \hspace{0.5cm} - 4 \mathrm{Im} [H Q^* + G Q^*] \Bigg[ p_+ . p_K (p_- . 
p_{K^*})^2 - p_- . p_K (p_+ . p_{K^*})^2 - p_- . p_+ p_K . p_{K^*} (p_- - p_+) . p_{K^*} \nonumber \\
&& \hspace{4.5cm} + (p_+ - p_-) . p_K \left( - m_{K^*}^2 p_- . p_+ + p_- . p_{K^*} p_+ . p_{K^*} \right) \Bigg] \nonumber \\
&& \hspace{0.5cm} - 4 \mathrm{Im} [Q N^*] \Bigg[ p_+ . p_{K^*} (p_- . p_{K})^2 - p_- . p_{K^*} (p_+ . p_{K})^2 - p_- . p_+ p_K . p_{K^*} (p_- - p_+) . p_{K} \nonumber \\
&& \hspace{3.0cm} + (p_+ - p_-) . p_{K^*} \left( p_- . p_{K} p_+ . p_{K} \right) \Bigg]
\nonumber \\
&& \hspace{0.5cm} + 4 |Q|^2 \Bigg[ - 2 p_- . p_K p_- . p_{K^*} p_+ . p_K p_+ . p_{K^*} + p_+ . p_- p_K . p_{K^*} (p_- - p_+) . p_K (p_- - p_+) . p_{K^*} 
\nonumber \\
&& \hspace{2.5cm} + (p_- . p_{K^*})^2 (p_+ . p_K)^2 + (p_+ . p_{K^*})^2 (p_- . p_K)^2 + m_{K^*}^2 p_- . p_K p_+ . p_K p_+ . p_- \nonumber \\
&& \hspace{2.5cm} - \left(\frac{1}{2} p_- . p_+\right)\left( m_{K^*}^2 (p_- . p_K)^2 + m_{K^*}^2 (p_+ . p_K)^2 \right) \Bigg] .
\eea
\noindent Note that in the above we have used the following constants:
\bea
A & = & 2\left( k_4 - k_1 k_3 \right) \left( V (C_9 - C_{10}) - 2 C_{7L} \left(\frac{i U_1}{m_B}\right) \frac{m_b}{q^2} \right) , \\
Q & = & 2\left( k_4 - k_1 k_3 \right) \left( V (C_9 + C_{10}) - 2 C_{7L} \left(\frac{i U_1}{m_B}\right) \frac{m_b}{q^2} \right) , \\
B & = & 2 \left( k_4 - k_1 k_3 \right) \left( C_{7L}\left(\frac{m_b U_2}{m_B q^2}\right) (m_B^2 - m_{K^*}^2 ) - A_1 (C_9 - C_{10}) \right) , \\
N & = & 2 \left( k_4 - k_1 k_3 \right) \left( C_{7L}\left(\frac{m_b U_2}{m_B q^2}\right) (m_B^2 - m_{K^*}^2 ) - A_1 (C_9 + C_{10}) \right) , \\
C & = & \left( \frac{- 2 k_1^2}{3 m_{K^*}^2} k_3 - 2 k_1 (k_4 - k_1 k_3) \right) \left( C_{7L}\left(\frac{m_b U_2}{m_B q^2}\right) (m_B^2 - m_{K^*}^2 ) - A_1 (C_9 - C_{10}) \right) \nonumber \\
&& + \left( 2 (k_4 - k_1 k_3 ) + \frac{2}{3} k_1^2 k_5 \right) \left( - \frac{1}{2} A_3 (C_9 - C_{10}) - C_{7L} \frac{m_b U_2}{m_B q^2} - \frac{2 U_3 m_b}{m_B q^2} C_{7L} \left( 1 + \frac{m_{K^*}^2 - p_b . p_{K^*}}{m_B^2 - m_{K^*}^2} \right) \right) , \nonumber \\
&& \\
G & = & \left( \frac{- 2 k_1^2}{3 m_{K^*}^2} k_3 - 2 k_1 (k_4 - k_1 k_3) \right) \left( C_{7L}\left(\frac{m_b U_2}{m_B q^2}\right) (m_B^2 - m_{K^*}^2 ) - A_1 (C_9 + C_{10}) \right) \nonumber \\
&& + \left( 2 (k_4 - k_1 k_3 ) + \frac{2}{3} k_1^2 k_5 \right) \left( - \frac{1}{2} A_3 (C_9 + C_{10}) - C_{7L} \frac{m_b U_2}{m_B q^2} - \frac{2 U_3 m_b}{m_B q^2} C_{7L} \left( 1 + \frac{m_{K^*}^2 - p_b . p_{K^*}}{m_B^2 - m_{K^*}^2} \right) \right) , \nonumber \\
&&  \\
D & = & \frac{2}{3} k_1^2 \left( C_{7L}\left(\frac{m_b U_2}{m_B q^2}\right) (m_B^2 - m_{K^*}^2 ) - A_1 (C_9 - C_{10}) \right) \nonumber \\
&& + \left(2 (k_4 - k_1 k_3)^2 + \frac{2}{3} k_1^2 k_5 \right) \left( - \frac{1}{2} A_2 (C_9 - C_{10}) - C_{7L} \frac{m_b U_2}{m_B q^2} - \frac{2 U_3 m_b}{m_B q^2} C_{7L} \left(\frac{m_{K^*}^2 - k_3}{m_B^2 - m_{K^*}^2} \right) \right) , \nonumber \\
&& \\
H & = & \frac{2}{3} k_1^2 \left( C_{7L}\left(\frac{m_b U_2}{m_B q^2}\right) (m_B^2 - m_{K^*}^2 ) - A_1 (C_9 + C_{10}) \right) \nonumber \\
&& + \left(2 (k_4 - k_1 k_3)^2 + \frac{2}{3} k_1^2 k_5 \right) \left( - \frac{1}{2} A_2 (C_9 + C_{10}) - C_{7L} \frac{m_b U_2}{m_B q^2} - \frac{2 U_3 m_b}{m_B q^2} C_{7L} \left(\frac{m_{K^*}^2 - k_3}{m_B^2 - m_{K^*}^2} \right) \right) , \nonumber \\
&&  
\eea
where $k_1 = \frac{1}{m_{K^*}^2}p_K . p_{K^*}$, $k_3 = p_{K^*} . p_b$, $k_4 = p_K . p_b$ and $k_5 = m_b^2 - \frac{k_3^2}{m_{K^*}^2}$.

\par If we now use the kinematics as prescribed in Kim \etal\cite{Kim:2000dq}, that is, where we set $p = \sqrt{p_{K^*}^2}$, $l = \sqrt{(p_+ + p_-)^2}$ and $\lambda = \frac{1}{4} ( m_B^2 - p^2 - l^2)^2 - p^2 l^2$. Furthermore, we shall introduce various angles, namely $\theta_K$ as the polar angle of the $K$ momentum in the rest frame of the $K^*$ meson with respect to the helicity axis, {\it i.e.} the outgoing direction of $K^*$. Similarly $\theta_+$ as the polar angle of the positron in the dilepton CM frame with respect to the $K^*$ momentum, and finally $\phi$ as the azimuthal angle between these planes, that is, the $K^* \to K \pi$ and $B \to K^* l^+ l^-$ planes. In this case our constants $k_i$ become:
$$ k_1 = \frac{p^2}{2 m_{K^*}^2} ~~,~~ k_3 = \frac{m_b}{m_B}\left(\sqrt{\lambda + p^2 l^2} + p^2 \right) ~~,~~ k_4 = \frac{k_3}{2} + \frac{m_b \sqrt{\lambda}}{2 m_B} \cos \theta_K ~~,$$
$$ \mathrm{and} \qquad k_5 = m_b^2 \left( 1 - \frac{\left(\sqrt{\lambda + p^2 l^2}\right)^2}{m_{K^*}^2 m_B^2} \right) . $$
In which case our previously defined constants will have the angular structure:
\bea
A = A_{(1)} + A_{(2)} \cos \theta_K &\hspace{20pt}& Q = Q_{(1)} + Q_{(2)} \cos \theta_K \nonumber \\
B = B_{(1)} + B_{(2)} \cos \theta_K && N = N_{(1)} + N_{(2)} \cos \theta_K \nonumber \\
C = C_{(1)} + C_{(2)} \cos \theta_K && G = G_{(1)} + G_{(2)} \cos \theta_K \nonumber \\
D = D_{(1)} + D_{(2)} \cos \theta_K + D_{(3)} \cos^2 \theta_K && H = H_{(1)} + H_{(2)} \cos \theta_K + H_{(3)} \cos^2 \theta_K , \nonumber
\eea
where we have defined these additional constants after the various integrations in appendix B.

\par The decay width for the full process can now be expressed as:
\bea
\frac{d \Gamma}{dp^2 dl^2 d(\cos\theta_K) d(\cos\theta_+) d\phi} & = & \frac{2 \sqrt{\lambda}}{128 \times 256 \pi^6 m_B^3} \left| \frac{\alpha G_F V_{tb} V_{ts}^* g}{\sqrt{2} \pi m_{K^*}} \frac{1}{ p^2 - m_{K^*}^2 + i m_{K^*} \Gamma_{K^*}} \right|^2 \left( |{\cal A}_L|^2 + |{\cal A}_R|^2 \right) . \nonumber \\
\eea
Note that as we have done the $p^2$ integration in the narrow resonance limit of $K^*$ 
\bea
\stackrel{\ds lim}{\Gamma_{K^*} \to 0} \frac{m_{K^*} \Gamma_{K^*}}{ (p^2 - m_{K^*}^2 )^2 + m_{K^*}^2 \Gamma_{K^*}^2} & = & \pi \delta (p^2 - m_{K^*}^2) , 
\eea
the double differential decay rate can be expressed as:
\bea
\frac{d \Gamma}{dl^2 d\phi} & = & \frac{ 15 \alpha^2 G_F^2 \left| V_{tb} V_{ts}^* \right|^2 m_{K^*}^7 }{ 256 \sqrt{2} \pi^5 m_B^3 \lambda^2} \int dp^2 d(\cos \theta_K) d(\cos \theta_+) \delta (p^2 - m_{K^*}^2) \left\{ \left| {\cal A}_L \right|^2 + \left| {\cal A}_R \right|^2 \right\} .
\eea
Note that from now on $l^2 = s$, the dilepton frame CM energy. Again, the constants $A_{(i)}$ etc, after the $p^2$ integration, have been defined in appendix B.

%%%%%%%%%%%%%%%%%%%%%%%%%%%%%%%%%%%%%%%%%%%%%%%%%%%%%%%%%%%%%%%%%%%%%
%
%  Section 3: Numerical results and discussion
%
%%%%%%%%%%%%%%%%%%%%%%%%%%%%%%%%%%%%%%%%%%%%%%%%%%%%%%%%%%%%%%%%%%%%%

\section{Numerical results and discusssion}

\par The parameters in the last equation for the differential decay rate are known and summarized in appendix C. The fivefold differential decay rate can easily be integrated. The integration over the variable $p^2$ is trivial in the zero width approximation as stated before. Integration over the polar angles $\theta_K$ and $\theta_+$ is easily done through Mathematica to yield the differential rate $\frac{d\Gamma}{ds d\phi}$. Figure 2 shows our results for this quantity which can be easily compared with experimental results as and when they become available. More important from the theoretical point of view is the dependence of this quantity on the angle $\phi$. This depends crucially on the nature of the Hamiltonian and also on the spin of the $K^*$ resonance which decays into hadrons. Following Kim \etal\cite{Kim:2000dq} we define the normalized differential rate:
\beq
r(\phi, s) =\frac{\ds \frac{d\Gamma}{ds d\phi}}{\ds \frac{d \Gamma}{ds}} ~.~
\eeq
We show in figure 3 a plot of this quantity, which once again is easily experimentally accessible once this process is seen. Our results as given in figures 2-5 show considerable structure and therefore would be a good test of the theory when comparing with experimental data. The results are similar to the parallel case for the similar processes with the  $K^*(892)$ resonances. However, the two involve different hadronic form factors and thus involve different combinations of the Wilson coefficients. Simultaneous comparison of the theoretical results for both these channels would thus be a more complete check of the underlying theory. The numerical results that we have presented are for values of Wilson coefficents calculated on the basis of the SM. In theories involving new physics, which would change the Wilson coefficents or add other coefficients (possibly scalar and pseudoscalar hadronic currents), our expressions are still valid. Note that the numerical results would change depending on the exact nature of the new physics introduced, and thus would again be a good situtation to confront theoretical results with experimental data. In summary, we have presented in this paper theoretical predictions based on the standard effective Hamiltonian for FCNC processes together with the LEET predictions of the hadronic form factors for the process $B \to K^*_2 l^+ l^-$, which we hope will soon be experimentally observed.

%%%%%%%%%%%%%%%%%%%%%%%%%%%%%%%%%%%%%%%%%%%%%%%%%%%%%%%%%%%%%%%%%%%%%
%
%  The Figures
%
%%%%%%%%%%%%%%%%%%%%%%%%%%%%%%%%%%%%%%%%%%%%%%%%%%%%%%%%%%%%%%%%%%%%%

\begin{figure}
\begin{center}
\epsfig{file=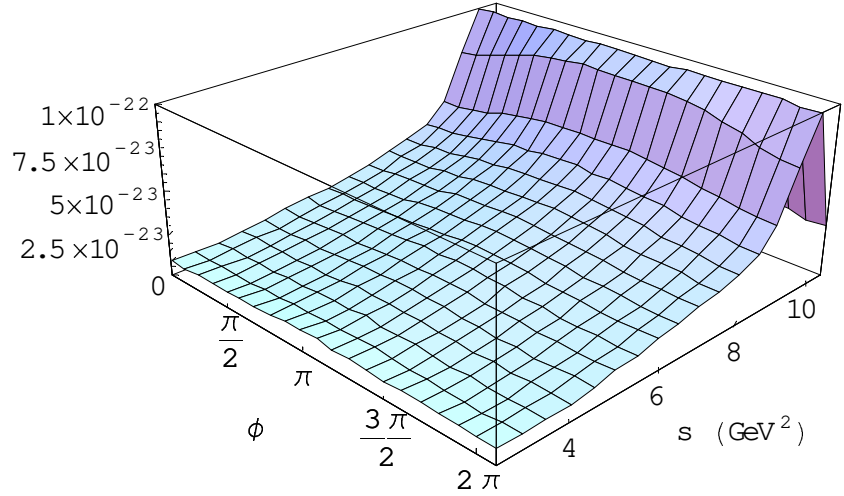,width=.62\textwidth}
\caption{\it The double differential decay rate $\frac{d \Gamma}{ds d\phi}$.}  
\label{fig:2}
\end{center}
\end{figure}

\begin{figure}
\begin{center}
\epsfig{file=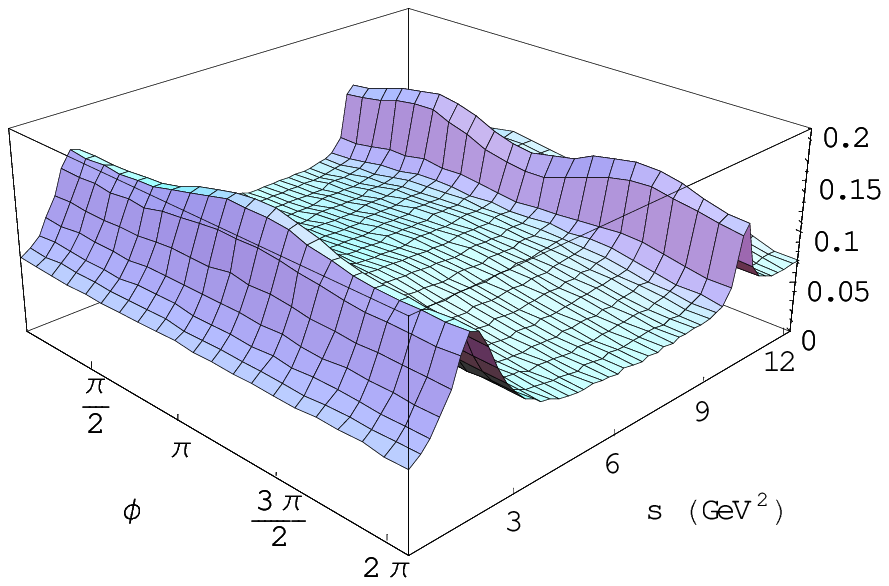,width=.62\textwidth}
\caption{\it The ratio of the double differential decay and the differential decay rate, $\frac{d^2 \Gamma}{ds d\phi}/\frac{d\Gamma}{ds}$.}  
\label{fig:3}
\end{center}
\end{figure}

\begin{figure}
\begin{center}
\epsfig{file=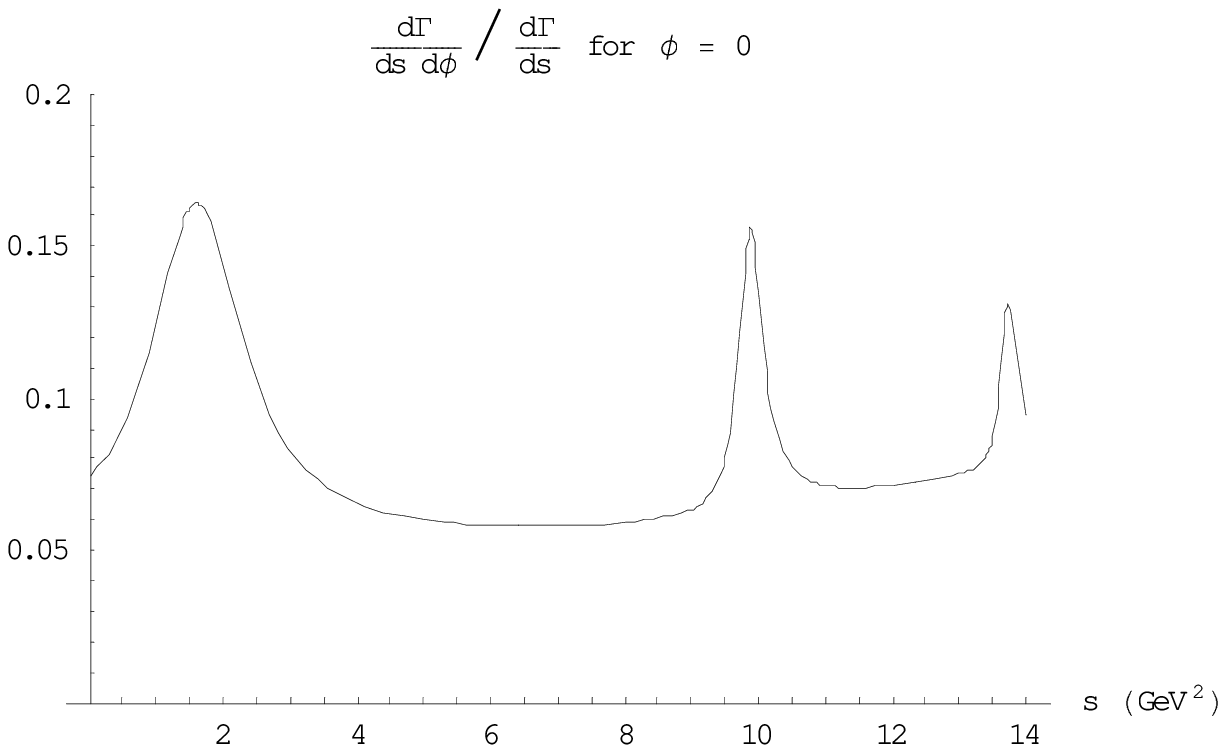,width=.62\textwidth}
\caption{\it The ratio of the double differential decay and the differential decay rate, $\frac{d^2 \Gamma}{ds d\phi}/\frac{d\Gamma}{ds}$, for $\phi = 0$.}  
\label{fig:4}
\end{center}
\end{figure}

\begin{figure}
\begin{center}
\epsfig{file=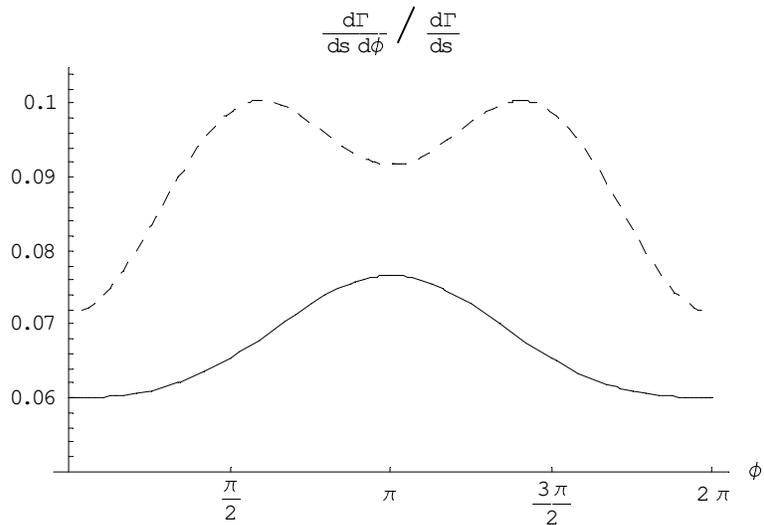,width=.65\textwidth}
\caption{\it The ratio of the double differential decay and the differential decay rate, $\frac{d^2 \Gamma}{ds d\phi}/\frac{d\Gamma}{ds}$, for $s = 5$GeV$^2$ (the solid line), and $s = 12$GeV$^2$ (the dashed line).}   
\label{fig:5}
\end{center}
\end{figure}

%%%%%%%%%%%%%%%%%%%%%%%%%%%%%%%%%%%%%%%%%%%%%%%%%%%%%%%%%%%%%%%%%%%%%
%
%  Acknowledgements
%
%%%%%%%%%%%%%%%%%%%%%%%%%%%%%%%%%%%%%%%%%%%%%%%%%%%%%%%%%%%%%%%%%%%%%

\section*{Acknowledgements}

The work of SRC was supported by the Department of Science \& Technology (DST), India under grant no SP/S2/K-20/99. The work of ASC was supported by the Japan Society for the Promotion of Science (JSPS), under fellowship no P04764.

%%%%%%%%%%%%%%%%%%%%%%%%%%%%%%%%%%%%%%%%%%%%%%%%%%%%%%%%%%%%%%%%%%%%%
%
%  The Appendices
%
%%%%%%%%%%%%%%%%%%%%%%%%%%%%%%%%%%%%%%%%%%%%%%%%%%%%%%%%%%%%%%%%%%%%%

\appendix

\section{The form factors}

\par The form factors we shall take from Cheng \etal \cite{Cheng:2004yj}, where the remaining form factors can be related to these using the relations in Charles \etal \cite{Dugan:1990de},  which lead to:
\begin{center}
$V = i U_1/m_B^2$ ~~,~~ $A_1 = - U_2/2$ ~~,~~ $A_2 = 0$ ~~,~~ $A_3 = 2 U_3/m_B^2$ .
\end{center}
Therefore:
\bea
U_1 (s) & = & \frac{0.19}{1 - 2.22 (s/m_B^2) + 2.13 (s/m_B^2)^2} , \nonumber \\
U_2 (s) & = & \frac{0.19}{\left( 1 - s/m_B^2 \right) \left( 1 - 1.77 (s/m_B^2) + 4.32 (s/m_B^2)^2 \right)} , \nonumber \\
U_3 (s) & = & \frac{0.16}{1 - 2.19 (s/m_B^2) + 1.80 (s/m_B^2)^2} . \nonumber
\eea

\section{The constants after $p^2$ integration}

\par After the $p^2$ integration the additional constants are defined as:
\bea
A_{(1)} & = & 0 , \\
A_{(2)} & = & \frac{m_b}{m_B} \sqrt{\lambda} \left( V (C_9 - C_{10}) - 2 C_{7L} \left( \frac{i U_1}{m_B} \right) \frac{m_b}{s} \right) , \\
Q_{(1)} & = & 0 , \\
Q_{(2)} & = & \frac{m_b}{m_B} \sqrt{\lambda} \left( V (C_9 + C_{10}) - 2 C_{7L} \left( \frac{i U_1}{m_B} \right) \frac{m_b}{s} \right) , \\
B_{(1)} & = & 0 , \\
B_{(2)} & = & \frac{m_b}{m_B} \sqrt{\lambda} \left( C_{7L} \left( \frac{m_b U_2}{m_B s} \right) (m_B^2 - m_{K^*}^2) - A_1 (C_9 - C_{10}) \right) , \\
N_{(1)} & = 0 , \\
N_{(2)} & = & \frac{m_b}{m_B} \sqrt{\lambda} \left( C_{7L} \left( \frac{m_b U_2}{m_B s} \right) (m_B^2 - m_{K^*}^2) - A_1 (C_9 + C_{10}) \right) , \\
C_{(1)} & = & - \frac{m_b}{6 m_{K^*}^2 m_B} \left( \sqrt{\lambda + m_{K^*}^2 s} + m_{K^*}^2 \right) \left( C_{7L} \left( \frac{m_b U_2}{m_B s} \right) (m_B^2 - m_{K^*}^2 ) - A_1 (C_9 - C_{10}) \right) \nonumber \\
&& + \frac{m_b^2}{6} \left( 1 - \frac{ \left( \sqrt{\lambda + m_{K^*}^2 s} + m_{K^*}^2 \right)}{ m_{K^*}^2 m_B^2 } \right) \nonumber \\
&& \times \left( - \frac{A_3}{2} (C_9 - C_{10}) - C_{7L} \frac{m_b U_2}{m_B s} - \frac{2 U_3 m_b}{m_B s} C_{7L} \left( 1 + \frac{m_{K^*}^2 - \frac{m_b}{m_B} \left( \sqrt{\lambda + m_{K^*}^2 s} + m_{K^*}^2 \right)}{m_B^2 - m_{K^*}^2} \right) \right) , \nonumber \\ \\
C_{(2)} & = & \frac{\sqrt{\lambda} m_b}{m_B} \Bigg[ - \frac{1}{2} \left( C_{7L} \left( \frac{m_b U_2}{m_B s} \right) ( m_B^2 - m_{K^*}^2 ) - A_1 (C_9 - C_{10}) \right) \nonumber \\
&& - \frac{A_3}{2} (C_9 - C_{10}) - C_{7L} \frac{m_b U_2}{m_B s} - \frac{2 U_3 m_b C_{7L}}{m_B s} \left( 1 + \frac{m_{K^*}^2 - \frac{m_b}{m_B} \left( \sqrt{\lambda + m_{K^*}^2 s} + m_{K^*}^2 \right) }{m_B^2 - m_{K^*}^2} \right) , \\
G_{(1)} & = & - \frac{m_b}{6 m_{K^*}^2 m_B} \left( \sqrt{\lambda m_{K^*}^2 s} + m_{K^*}^2 \right) \left( C_{7L} \left( \frac{m_b U_2}{m_B s} \right) (m_B^2 - m_{K^*}^2 ) - A_1 (C_9 + C_{10}) \right) \nonumber \\
&& + \frac{m_b^2}{6} \left( 1 - \frac{ \left( \sqrt{\lambda + m_{K^*}^2 s} + m_{K^*}^2 \right)}{ m_{K^*}^2 m_B^2 } \right) \nonumber \\
&& \times \left( - \frac{A_3}{2} (C_9 + C_{10}) - C_{7L} \frac{m_b U_2}{m_B s} - \frac{2 U_3 m_b}{m_B s} C_{7L} \left( 1 + \frac{m_{K^*}^2 - \frac{m_b}{m_B} \left( \sqrt{\lambda + m_{K^*}^2 s} + m_{K^*}^2 \right)}{m_B^2 - m_{K^*}^2} \right) \right) , \nonumber \\ \\
G_{(2)} & = & \frac{\sqrt{\lambda} m_b}{m_B} \Bigg[ - \frac{1}{2} \left( C_{7L} \left( \frac{m_b U_2}{m_B s} \right) ( m_B^2 - m_{K^*}^2 ) - A_1 (C_9 + C_{10}) \right) \nonumber \\
&& - \frac{A_3}{2} (C_9 + C_{10}) - C_{7L} \frac{m_b U_2}{m_B s} - \frac{2 U_3 m_b C_{7L}}{m_B s} \left( 1 + \frac{m_{K^*}^2 - \frac{m_b}{m_B} \left( \sqrt{\lambda + m_{K^*}^2 s} + m_{K^*}^2 \right)}{m_B^2 - m_{K^*}^2} \right) \Bigg] , \\
D_{(1)} & = & \frac{1}{6} \left( C_{7L} \frac{m_b U_2}{m_B s} (m_B^2 - m_{K^*}^2) - A_1 (C_9 - C_{10}) \right) + \frac{m_b^2}{6} \left( 1 - \frac{\left(\sqrt{\lambda + m_{K^*}^2 s} + m_{K^*}^2 \right)^2}{m_{K^*}^2 m_B^2} \right) \nonumber \\
&& \times \left( - \frac{A_2}{2} (C_9 - C_{10}) - C_{7L} \frac{m_b U_2}{m_B s} - \frac{2 U_3 m_b C_{7L}}{m_B s} \left( \frac{m_{K^*}^2 - \frac{m_b}{m_B} \left( \sqrt{\lambda + m_{K^*}^2 s} + m_{K^*}^2 \right)}{m_B^2 - m_{K^*}^2} \right) \right), \\
D_{(2)} & = & 0 , \\
D_{(3)} & = & \frac{m_b^2 \lambda}{2 m_B^2} \left( - \frac{A_2}{2} (C_9 - C_{10}) - C_{7L} \frac{m_b U_2}{m_B s} - \frac{2 U_3 m_b C_{7L}}{m_B s} \left( \frac{m_{K^*}^2 - \frac{m_b}{m_B} \left( \sqrt{\lambda + m_{K^*}^2 s} + m_{K^*}^2 \right)}{m_B^2 - m_{K^*}^2} \right) \right) , \nonumber \\ \\
H_{(1)} & = & \frac{1}{6} \left( C_{7L} \frac{m_b U_2}{m_B s} (m_B^2 - m_{K^*}^2) - A_1 (C_9 + C_{10}) \right) + \frac{m_b^2}{6} \left( 1 - \frac{\left(\sqrt{\lambda + m_{K^*}^2 s} + m_{K^*}^2 \right)^2}{m_{K^*}^2 m_B^2} \right) \nonumber \\
&& \times \left( - \frac{A_2}{2} (C_9 + C_{10}) - C_{7L} \frac{m_b U_2}{m_B s} - \frac{2 U_3 m_b C_{7L}}{m_B s} \left( \frac{m_{K^*}^2 - \frac{m_b}{m_B} \left( \sqrt{\lambda + m_{K^*}^2 s} + m_{K^*}^2 \right)}{m_B^2 - m_{K^*}^2} \right) \right) , \\
H_{(2)} & = & 0 , \\
H_{(3)} & = & \frac{m_b^2 \lambda}{2 m_B^2} \left( - \frac{A_2}{2} (C_9 + C_{10}) - C_{7L} \frac{m_b U_2}{m_B s} - \frac{2 U_3 m_b C_{7L}}{m_B s} \left( \frac{m_{K^*}^2 - \frac{m_b}{m_B} \left( \sqrt{\lambda + m_{K^*}^2 s} + m_{K^*}^2 \right)}{m_B^2 - m_{K^*}^2} \right) \right) . \nonumber \\
\eea

\section{Input parameters and Wilson coefficients\label{appendix:b}}

\par The input parameters used in the generation of the numerical results are as follows \cite{Yao:2006px}:
\begin{center}
$m_B = 5.26$GeV ~~,~~ $m_{K^*} = 1.43$GeV ~~,~~ $m_b = 4.8$GeV ~~,~~ $m_c = 1.4$GeV , \\
$m_s = 0.1$GeV ~~,~~ ${\cal B}(J/\psi(1S) \to \ell^+ \ell^-) = 6 \times 10^{-2}$ , \\ $m_{J/\psi(1S)} = 3.097$GeV ~~,~~ ${\cal B}(\psi(2S) \to \ell^+ \ell^-) = 7.3 \times 10^{-3}$ , \\$m_{\psi(2S)} = 3.686$GeV ~~,~~ $\Gamma_{\psi(2S)} = 0.277 \times 10^{-3}$GeV , \\ $\Gamma_{J/\psi(1S)} = 0.093 \times 10^{-3}$GeV ~~,~~ $V_{tb} V_{ts}^* = 0.0385$ ~~,~~ $\alpha = \frac{1}{129}$ ~~,~~ $G_F = 1.17 \times 10^{-5}$ GeV$^{-2}$.
\end{center}

\noindent The Wilson coefficients used were as in Kim \etal \cite{Kim:2000dq}, namely:
\bea
C_{7L} = - 0.285 & , & C_{10} = - 4.546 , \nonumber
\eea
\bea
C_9 & = & 4.153 + 0.381 g\left(\frac{m_c}{m_b} , \frac{s}{m_B^2} \right) + 0.033 g\left(1, \frac{s}{m_B^2} \right) + 0.032 g\left(0, \frac{s}{m_B^2} \right) - 0.381 \times 2.3 \times \frac{3 \pi}{\alpha} \nonumber \\
&& \hspace{1 cm} \times \left( \frac{\Gamma_{\psi(2S)} {\cal B}(\psi(2S) \to \ell^+ \ell^-) m_{\psi(2S)}}{s - m_{\psi(2S)}^2 + i m_{\psi(2S)} \Gamma_{\psi(2S)}} + \frac{\Gamma_{J/\psi(1S)} {\cal B}(J/\psi(1S) \to \ell^+ \ell^-) m_{J/\psi(1S)}}{s - m_{J/\psi(1S)}^2 + i m_{J/\psi(1S)} \Gamma_{J/\psi(1S)}}\right) , \nonumber
\eea
where the function $g$ is taken from reference \cite{gfunc}:
\bea
g(\hat{m}_i, \hat{s}) & = & - \frac{8}{9} \ln (\hat{m}_i) + \frac{8}{27} + \frac{4}{9} \left( \frac{4 \hat{m}_i^2}{\hat{s}} \right) - \frac{2}{9} \left( 2 + \frac{4 \hat{m}_i^2}{\hat{s}} \right) \sqrt{ \left| 1 - \frac{4 \hat{m}_i^2}{\hat{s}} \right|} \nonumber \\
&& \hspace{1 in} \times \left\{ \begin{array}{lc}
\left| \ln \left(\frac{1 + \sqrt{1 - 4 \hat{m}_i^2/\hat{s}}}{1 - \sqrt{1 - 4 \hat{m}_i^2/\hat{s}}} \right) - i \pi \right| & , 4 \hat{m}_i^2 < \hat{s} \\
2 \arctan \frac{1}{\sqrt{4 \hat{m}_i^2/\hat{s} - 1}} & , 4 \hat{m}_i^2 > \hat{s}
\end{array} \right. \nonumber
\eea

\end{document}